\documentclass[12pt]{article}
\usepackage{amsmath}
\usepackage{amsfonts}
\usepackage{amssymb}
\usepackage{graphicx}
\usepackage{latexsym,amsmath,amssymb,amsfonts,amsthm,enumerate}

\gdef\ffrac#1#2{\textstyle{#1\over#2}\displaystyle}
\gdef\refcite{\cite}
\begin{document}
\setcounter{page}{0} \topmargin 0pt
\renewcommand{\thefootnote}{\arabic{footnote}}
\newpage
\setcounter{page}{0}

\begin{titlepage}

\begin{center}
{\Large {\bf Quantum Network Models and Classical Localization Problems\footnote{To
appear in {\sl Fifty years of Anderson Localization}, E.~Abrahams, ed. (World Scientific).}}}\\

\vspace{2cm} {\large John Cardy$^{a,b}$\\}  \vspace{0.5cm} {\em
$^{a}$Rudolf Peierls Centre for Theoretical Physics\\ 1 Keble
Road, Oxford OX1 3NP, UK\\}  \vspace{0.2cm} {\em $^{b}$All Souls
College, Oxford\\}

\vspace{2cm}

March 2010

\end{center}

\vspace{1cm}

\begin{abstract}
A review is given of quantum network models in class C which, on a
suitable 2d lattice, describe the spin quantum Hall plateau
transition. On a general class of graphs, however, many
observables of such models can be mapped to those of a classical
walk in a random environment, thus relating questions of quantum
and classical localization. In many cases it is  possible to make
rigorous statements about the latter through the relation to
associated percolation problems, in both two and three dimensions.

\end{abstract}

\end{titlepage}

\section{Introduction}\label{sec:intro}

Lattice models of spatially extended systems have a long record of
usefulness in condensed matter physics. Even when the microscopic
physics is not necessarily related to a crystalline lattice, it
can be very useful to concentrate the essential degrees of freedom
onto a regular lattice whose length scale is larger than the
microscopic one yet much smaller than that the expected scale of
the physical phenomena the model is designed to address. In many
cases, the phenomenon of universality ensures that this
idealization can nevertheless reproduce certain aspects exactly.
The classic example is that of a lattice gas, where a
coarse-grained lattice on the scale of the particle interaction
radius is introduced and used to make predictions for continuum
systems, in cases in which the correlation length is large, for
example close to the liquid-gas critical point.

The lattice models discussed in this article -- in this context
called network models -- were first introduced by Chalker and
Coddington\cite{ChalkerCodd} as a theoretical model for
non-interacting electrons in two dimensions in a strong transverse
magnetic field and in the presence of disorder: the physical
setting for the integer quantum Hall effect. The starting point is
to consider non-interacting electrons moving in two dimensions in
a disordered potential $V(r)$ and a strong perpendicular magnetic
field $B$. We assume that the length scale of variation of $V(r)$
is much larger than the magnetic length. In this limit the
electronic motion has two components with widely separated time
scales\cite{PrangeJoynt}: the cyclotron motion and the motion of
the guiding center, along contours of $V(r)$. The total energy of
the electron in this approximation is
$E=(n+\frac12)\hbar\omega_c+V(r)$, where $\omega_c=eB/m$ and $n$
labels the Landau levels, and we therefore expect to find extended
states at energies $E_c$ corresponding to those values of $V$ at
which the contours of $V(r)$ percolate. In 2d this is expected to
occur at one particular value of $V$, which can be taken to be
$V=0$. Otherwise, away from the percolation threshold the guiding
centers are confined to the neighborhoods of the closed contours,
corresponding to bulk insulating phases which conduct only along
the edges of the sample. This immediately provides a simple
explanation of the experimental result that extended states occur
only at the transition and not in the Hall plateaux. If this were
literally correct the plateau transition would be in the same
universality class as classical percolation.
\begin{figure}
\centerline{\includegraphics[width=5cm]{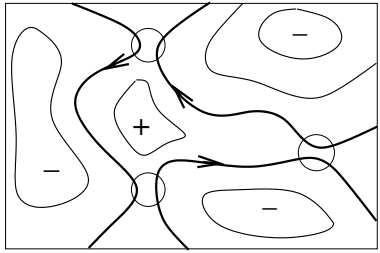}}
\caption{ A typical configuration of contours of the random
potential $V(r)$. In the guiding center approximation the particle
follows these. Only those with $V\approx0$ (thicker lines) are
important for the plateau transition. Quantum tunnelling can occur
close to the saddle points of $V(r)$ (circled). Figure adapted
from Ref.~\refcite{ChalkerCodd}.} \label{fig:contours}
\end{figure}
\begin{figure}
\centerline{\includegraphics[width=5cm]{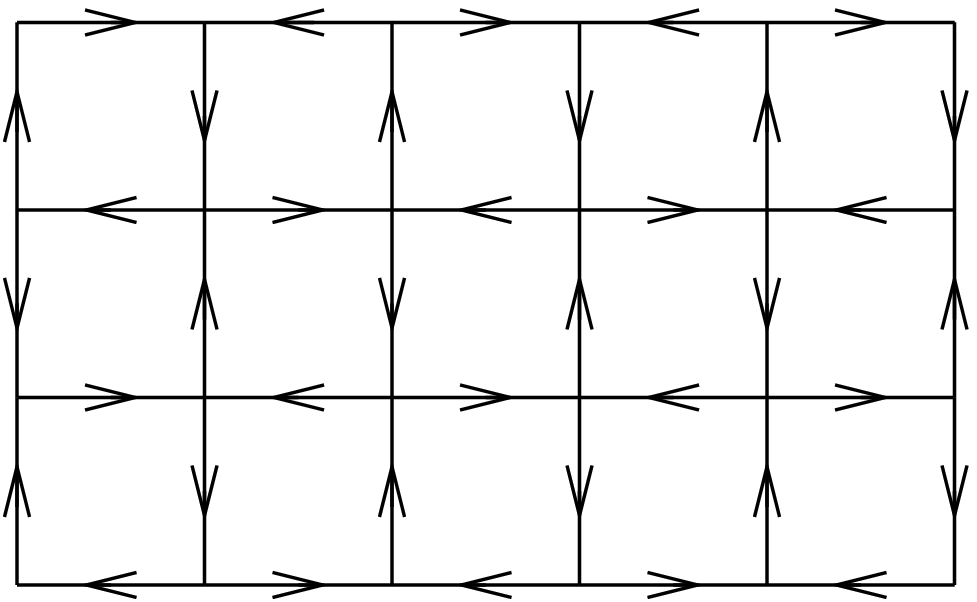}}
\caption{The idealization of Fig.~\ref{fig:contours} on the
L-lattice. The edges correspond to the contours with $V\approx0$
and tunnelling occurs at the nodes.} \label{fig:Llattice}
\end{figure}

However, this picture is modified for energies close to the
transition, since quantum tunnelling is expected to be
important\cite{Trugman} where closed contours approach each other,
see Fig.~\ref{fig:contours}. The network model idealizes this
picture by distorting the percolating contours of $V(r)$ into a
regular square lattice, known as the L-lattice, shown in
Fig.~\ref{fig:Llattice}. In this approximation the potential $V$
takes a checkerboard form, being $>0$ on (say) even squares and
$<0$ on odd squares. In the limit of large magnetic field the spin
degree of freedom of the electrons can be ignored and there is a
one-dimensional vector space associated with each edge. The
quantum tunnelling at each node is taken into account by an
$S$-matrix which connects the spaces on the adjacent incoming and
outgoing edges. This depends on $E$ in such a way that for $E>E_c$
tunnelling between edges bordering regions with $V>0$ is enhanced,
and \em vica versa \em for $E<E_c$. Apart from this, the most
important quantum feature of the problem is the phase which the
electron wave function picks up on traversing a closed contour,
which is proportional to the magnetic flux through the loop, and
therefore its area. On idealizing the loops to a regular lattice,
this is represented by a quenched uncorrelated random flux through
each plaquette, or, equivalently, a quenched random U$(1)$ phase
as the particle traverses a given edge. While in principle the
connectivity of the lattice, the $S$-matrix elements at the nodes,
and the U$(1)$ phases on the edges are all quenched random
variables, in fact only the latter appear to be the most relevant
in describing the universal properties of the transition.

The Chalker-Coddington model was initially analyzed
numerically\cite{ChalkerCodd} using transfer matrix methods. Its
predictions appear to agree remarkably well with experimental
results, perhaps embarrassingly so since it ignores
electron-electron interactions which may become important near the
transition. However it has so far resisted all attempts at an
analytic solution (as have other more sophisticated field
theoretic formulations of the integer quantum Hall plateau
transition\cite{fieldtheories}), which by now perhaps elevates
this to being one of the outstanding unsolved problems of
mathematical physics.

Later, following interest in various forms of exotic
superconductivity, it was suggested that certain disordered
spin-singlet superconductors, in which time-reversal symmetry is
broken for orbital motion but Zeeman splitting is negligible,
should exhibit a quantum spin Hall effect, in which the role of
the electric current is replaced by that of a spin
current.\cite{spinHall} The single-particle hamiltonians for such
a system then turn out to possess an Sp$(2)$ (or equivalently
SU$(2)$) symmetry. In the classification scheme of localization
universality classes due to Altland and Zirnbauer\cite{AZ} they
are labelled as class C. The corresponding variant of the
Chalker-Coddington model is straightforward to write down, and was
studied, once again numerically\cite{ClassCnum,spinHall}. Then, in
a remarkable paper, Gruzberg, Read and Ludwig\cite{GRL} argued
that several important ensemble-averaged properties of this model
(including the conductance) are simply related to those of
critical \em classical \em 2d percolation. This is a powerful
result because many of the universal properties of percolation are
known rigorously\cite{SLE}. It therefore gives exact information
about a non-trivial quantum localization transition.

The arguments of Gruzberg \em et al. \em\cite{GRL} were based on a
transfer matrix formulation of the problem and therefore
restricted to the particular oriented lattice (the L-lattice) used
by the Chalker-Coddington model, which is appropriate to the
quantum Hall problem in 2d. They also used supersymmetry to
perform the quenched average. Subsequently, Beamond, Cardy and
Chalker\cite{BCC} gave an elementary, albeit long-winded, proof of
their main result which holds for any lattice of coordination
number four, and any orientation of this lattice as long each node
has two incoming and two outgoing edges. Later, this was shown in
a slightly more elegant fashion using supersymmetry\cite{JC}. In
each case certain quenched averages of the quantum problem are
related to observables of a certain kind of \em classical \em
random walk on the same lattice. If the quantum states are
localized, the corresponding classical walks close after a finite
number of steps. If the quantum states are extended, in the
classical problem the walks can escape to infinity.

Since this correspondence holds on a very general set of graphs
and lattices, it can be used more generally to improve our
understanding of quantum localization problems. In particular it
can test the generally accepted notion that in two dimensions all
states are localized, except in certain cases with special
symmetries (such as at the Hall plateau transition). It can help
understand why in higher dimensions there should be in general a
transition between localized and extended states, and possibly
illumine the nature of that transition. More mathematically, it
may shed light on the search for a rigorous proof of the existence
of extended states. Apart from the L-lattice considered by
Gruzberg \em et al.\em, it is possible in several cases to use
known information about percolation to place bounds on the
behavior of the classical walks and hence the quantum network
model. Up until recently, these arguments have been restricted to
two dimensions, but now suitable three-dimensional lattices have
been identified in which the correspondence to classical
percolation is explicit.

The layout of this paper is as follows. In Sec.~\ref{sec:obs} we
describe general network models and observables which are related
to experimentally measurable quantities. In Sec.~\ref{sec:proof}
we summarize the supersymmetric proof\cite{JC} of the main theorem
which relates suitable quenched averages of these observables in
the Sp$(2)$ network model on a general graph to averages in a
classical random walk problem on the same graph. The next section
\ref{sec:2d} describes how these classical models on certain
lattices (the L-lattice, relevant to the quantum Hall effect, and
the Manhattan lattice) relate to 2d classical percolation which
can then be used to bound their behavior. In Sec.~\ref{sec:3d} we
extend this to some special 3d lattices, describing relatively
recent work, some of it so far unpublished. Finally in
Sec.~\ref{sec:further} we discuss some outstanding problems.

\section{General network models}\label{sec:obs}
\begin{figure}
\centerline{\includegraphics[width=3.5cm]{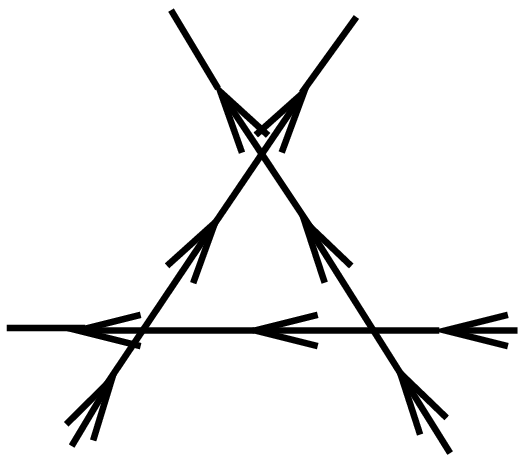}}
\caption{A $3\to3$ node which can be decomposed into $2\to2$
nodes.} \label{fig:3to3}
\end{figure}

In this section we define a general network model on a graph $\cal
G$ and discuss the kind of observables we would like to calculate.
The graph $\cal G$ consists of nodes $n$ and oriented edges.
Initially suppose that $\cal G$ is closed, that is every edge
connects two nodes, and that each node has exactly two incoming
and two outgoing edges. In fact the general theorem to be proved
in Sec.~\ref{sec:proof} holds for more general graphs, but it can
be shown\cite{JC} that the corresponding classical problem has
non-negative weights (and so admits a probabilistic
interpretation) if and only if each node in $\cal G$ and its
correspond transition amplitudes can be decomposed into a skeleton
graph with only $2\to2$ nodes. See Fig.~\ref{fig:3to3}.

On each edge $e$ of $\cal G$ is an $N$-dimensional Hilbert space
${\cal H}_e$. We assume these are all isomorphic. The Hilbert
space of the whole system is then $\otimes_{e\in{\cal G}}{\cal
H}_e$. We consider a single particle whose wave function at at
time $t$ is a superposition of the basis states in this space. The
dynamics is discrete: if the particle is at the center of edge $e$
at time $t$, at the next time $t+1$ it must move in the direction
of the orientated edges through a node to the center of a
neighboring edge $e'$. Because of the discrete dynamics we should
consider the unitary time-evolution operator $\cal U$. This has an
off-diagonal block elements ${\cal U}_{e'e}$ which have the form
\[
{\cal U}_{e'e}=U_{e'}^{1/2}\,S_{e'e}\,U_e^{1/2}\,.
\]
Here $U_e$ is a unitary $N\times N$ matrix which maps ${\cal
H}_e\to{\cal H}_e$ and describes the rotation of the wave function
in the internal space as the particle moves along the edge $e$,
and $S_{e'e}$ maps ${\cal H}_e\to{\cal H}_{e'}$ and describes the
transmission through a node. The evolution matrix after time $t$
therefore has non-zero block elements
\[
{\cal
U}^t_{e_f,e_i}=\sum_{\gamma(e_t,e_0)}\delta_{e_t,e_f}\delta_{e_0,e_i}
U_{e_t}^{1/2}S_{e_t,e_{t-1}}U_{e_{t-1}}\cdots
U_{e_1}S_{e_1,e_0}U_{e_0}^{1/2}\,,
\]
where the sum is over all Feynman paths $\gamma(e_t,e_0)$ on $\cal
G$ of length $t$, starting at $e_i$ and ending at $e_f$. Note that
in this sum a given edge can be traversed an arbitrary number of
times.

Both the matrices $U_e$ and $S_{e'e}$ are quenched random
variables, assumed chosen from the invariant measure on some
subgroup of U$(N)$. This is chosen according to the symmetry class
under consideration. For spinless, or spin-polarized, electrons,
where electric charge is conserved, we can take $N=1$ and the
$U_e\in {\rm U}(1)$. For models with class C symmetry\cite{AZ},
corresponding for example to the spin quantum Hall effect, the
single-particle Hilbert space is even-dimensional, and there is an
action of $\sigma_y$ such that the single-particle hamiltonian
$\cal H$ satisfies ${\cal H}^*=-\sigma_y{\cal H}\sigma_y$. This
implies a symmetry between states with energies $\pm E$, and that
the time-evolution operator ${\cal U}^t=e^{-i{\cal H}t}$ satisfies
\[
{\cal U}^*=\sigma_y{\cal U}\sigma_y\,,
\]
implying that the matrices $U_e$ should be symplectic, in Sp$(N)$,
which for $N=2$ is isomorphic to SU$(2)$.

For a given node $n$ with incoming edges $(e_1,e_2)$ and outgoing
edges $(e_1',e_2')$, the $S$-matrix has the block form
\begin{equation}\label{See}
\left(\begin{array}{cc}S_{e_1'e_1}&S_{e_2'e_1}\\S_{e_1'e_2}&S_{e_2'e_2}
\end{array}\right)\,,
\end{equation}
where each element is an $N\times N$ matrix. However, since all
the matrices are chosen at random, there is the gauge freedom of
redefining $S_{e'e}\to V_{e'}^{-1}S_{e'e}V_e$, $U_e\to
V_e^{-1}U_eV_e$, which allows us to choose each $S_{e'e}$ to be
proportional to the unit $N\times N$ matrix ${\bf 1}_N$. The
remaining $2\times2$ matrix can further be chosen to be real and
therefore orthogonal. Thus in fact (\ref{See}) can be replaced by
\begin{equation}\label{cossin}
{\bf 1}_N\otimes\left(\begin{array}{cc}\cos\theta_n&\sin\theta_n\\
-\sin\theta_n&\cos\theta_n\end{array}\right)\,.
\end{equation}
We are left with the gauge-transformed $U_e$ and the $\theta_n$ as
quenched random variables. However, we shall treat them
differently, first keeping the $\theta_n$ fixed while performing
quenched averages over the $U_e$. In fact we shall see that in
most cases is suffices to take all the $\theta_n$ as fixed and
equal on each sublattice.

Because we consider discrete rather than continuous time
evolution, the usual Green function is replaced by the resolvent
$(1-z{\cal U})^{-1}$ of the unitary evolution operator, whose
matrix elements we shall however continue to refer to as the Green
function:
\begin{equation}\label{G1}
G(e_f,e_i;z)\equiv\langle e'|(1-z{\cal U})^{-1}|e\rangle\,.
\end{equation}
Here $|e\rangle\in{\cal H}$ has non-zero components only in ${\cal
H}_e$. Note that $G(e_f,e_i;z)$ is an $N\times N$ matrix mapping
${\cal H}_{e_i}\to{\cal H}_{e_f}$. The parameter $z$ is the analog
of the energy (roughly $z\sim e^{iE}$). For $|z|\ll1$, the
expansion of (\ref{G1}) in powers of $z$ gives $G$ as a sum over
Feynman paths from $e_i$ to $e_f$. Each path $\gamma$ is weighted
by $z^{|\gamma|}$ times an ordered product of the $U_e$ with
$e\in\gamma$ and the factors of $\cos\theta_n$ or
$\pm\sin\theta_n$ for $n\in\gamma$. For a finite closed graph
$\cal G$, this expansion is convergent for $|z|<1$ and therefore
defines $G$ as an analytic function in this region. In general,
for a finite $\cal G$, there are poles on the circle $|z|=1$
corresponding to the eigenvalues of $\cal U$.

However, for $|z|>1$, $G$ admits an alternative expansion in
powers of $z^{-1}$ by writing it as
\begin{equation}\label{G2}
G(e_f,e_i;z)\equiv-\langle e'|z^{-1}{\cal U}^\dag(1-z^{-1}{\cal
U}^\dag)^{-1}|e\rangle\,.
\end{equation}
This is given by a sum over paths $\gamma$ with length $\geq1$ of
a product of $z^{-|\gamma|}$ with ordered factors $U_e^\dag$ along
the path, and $\cos\theta_n$ or $\pm\sin\theta_n$ as before.

The eigenvalues of $\cal U$ have the form $e^{i\epsilon_j}$, where
the $-\pi<\epsilon_j\leq\pi$ with $j=1,\ldots,{\cal N}$ are
discrete for a finite graph $\cal G$. We define the \em density of
states \em by
\[
\rho(\epsilon)\equiv\frac1{\cal
N}\sum_j\delta(\epsilon-\epsilon_j)\,.
\]
In the standard way, the density of states is given by the
discontinuity in the trace of the Green function, this time across
$|z|=1$ rather than ${\rm Im}\,E=0$:
\begin{equation}\label{rho}
\rho(\epsilon)=\frac1{2\pi N|{\cal G}|}\sum_{e\in{\cal
G}}\lim_{\eta\to0+}\left({\rm Tr}\,G(e,e;z=e^{i\epsilon-\eta})-
{\rm Tr}\,G(e,e;z=e^{i\epsilon+\eta})\right)\,,
\end{equation}
where the trace is in the $N$-dimensional space ${\cal H}_e$. In
the case where $\cal G$ is a regular lattice, in the thermodynamic
limit we expect the eigenvalues to be continuously distributed
around the unit circle.

We note that in the U$(1)$ case, when the $U_e$ are pure phases
$e^{i\phi_e}$, in each term in the Feynman path expansion of
(\ref{G1},\ref{G2}) a given edge $e$ occurs with a weight
$e^{in^\gamma_e\phi_e}$, where $n^\gamma_e$ is the number of times
the path $\gamma$ traverses this edge. On averaging a given path
$\gamma$ will contribute to the mean density of states only if
$n^\gamma_e=0$, that is it has length zero. Thus, in the U$(1)$
network models, $\overline{G(e,e,z)}=1$ for $|z|<1$, and zero for
$|z|>1$, and the mean density of states is constant, and
completely independent of the $\theta_n$. This is consistent with
the general result that at the plateau transition in the charge
quantum Hall effect, the density of states is non-singular.

We now turn to the conductance. In order to define this, we must
consider an open graph, which can be obtained from a given closed
graph $\cal G$ by breaking open a subset $\{e\}$ of the edges,
relabelling each broken edge $e$ as $e_{\rm in}$ and $e_{\rm
out}$. External contacts are subsets $C_{\rm in}$ and $C_{\rm
out}$ of these. The transmission matrix $T$ is a rectangular
matrix with elements
\[
T=\langle e_{\rm out}|(1-{\cal U})^{-1}|e_{\rm in}\rangle\,,
\]
where $e_{\rm in}\in C_{\rm in}$, $e_{\rm out}\in C_{\rm out}$.
Note that for an open graph the resolvent $(1-z{\cal U})^{-1}$
generally has poles inside the unit circle $|z|=1$. In the
thermodynamic limit, however, as long as the fraction of broken
edges is zero (for example if we have contacts only along part of
the edge of the sample), it is believed that the limit as
$|z|\to1$ can be taken.

The multi-channel Landauer formula then gives the conductance
between the contacts as
\[
g=(Q^2/h){\rm Tr}\,T^\dag T\,,
\]
where $Q$ is the quantum of charge carried by the particle. For
the integer quantum Hall effect, $Q=e$, and for the spin Hall
effect $Q=\frac12\hbar$.

\section{The main theorems.}\label{sec:proof}
In this section we focus on the Sp$(2)$ (=SU$(2)$) case and
summarize the method of proof of the main theorem relating the
quenched averages of the density of states and conductances to
observables of a classical random walk model on the same graph
$\cal G$. We restrict attention to graphs with exactly 2 incoming
and 2 outgoing edges. The general case is considered in
Ref.~\refcite{JC} and is considerably more verbose.

\begin{figure}
\centerline{\includegraphics[width=6.5cm]{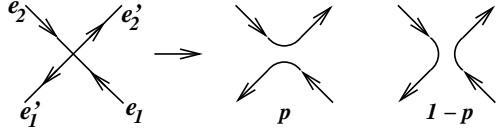}}
\caption{The two ways to decompose a node of $\cal G$. Note that
in general this does not have to be planar, although for the
L-lattice (Fig.~\ref{fig:Llattice}) it is.}  \label{fig:decomp}
\end{figure}

Let us first define the corresponding classical problem. Starting
with a given closed graph $\cal G$, to each node with incoming
edges $(e_1,e_2)$ and outgoing edges $(e_1',e_2')$ associate its
two possible \em decompositions \em $((e_1',e_1),(e_2',e_2))$ and
$((e_2',e_1),(e_1',e_2))$, corresponding to the two distinct ways
of passing twice through the node without using a given edge more
than once, irrespective of the order in which the edges are used.
This is illustrated in Fig.~\ref{fig:decomp}.
To each decomposition assign a probability $p_n=\cos^2\theta_n$ or
$1-p_n=\sin^2\theta_n$, corresponding to the $S$-matrix in
(\ref{cossin}). Doing this for each node in $\cal G$ gives a
decomposition of $\cal G$ into a union of closed loops. There are
$2^{|{\cal G}|}$ such decompositions, where $|{\cal G}|$ is the
total number of nodes, and the assigned probabilities give a
product measure on the set of decompositions.

We now state the two main theorems. We use the symbol $\overline
A$ to denote a quenched average in the original quantum network
model. Let $P(L,e)$ be the probability that the edge $e$ lies on a
closed loop of length $L$ in the classical problem.

\noindent{\bf Theorem 1.}\label{th1}\em $\overline{G(e_f,e_i,z)}$
vanishes unless $e_f=e_i$, when it is given by
\[
{\rm
Tr}\,\overline{G(e,e,z)}=\left\{\begin{array}{cc}2-\sum_{L>0}P(L,e)z^{2L}&:|z|<1\\
\sum_{L>0}P(L,e)z^{-2L}&:|z|>1\end{array}\right.\,.
\]
\em

If we apply this to the density of states in (\ref{rho}) we find
simply
\begin{equation}\label{dos}
\rho(\epsilon)=(1/2\pi)\big(1-\sum_{L>0}P(L)\cos(2L\epsilon)\big)\,,
\end{equation}
where $P(L)=(|{\cal G}|)^{-1}\sum_eP(L,e)$, the probability that a
edge chosen at random lies on a loop of length $L$. Note that if
this distribution decays sufficiently fast for large $L$,
$\rho(\epsilon)$ is expected to be analytic, while if it decays as
a power law, $\rho(\epsilon)$ will have a power law singularity at
$\epsilon=0$.

For an open graph with external contacts $C_{\rm in}$ and $C_{\rm
out}$, the decomposition of $\cal G$ can also lead to open paths
connecting edges in $C_{\rm in}$ to those in $C_{\rm out}$. In
this case, for the conductance, we have

\noindent{\bf Theorem 2.}\label{th2} \em The mean conductance is
\[
\overline g=2\sum_{e_{\rm in}\in C_{\rm in}}\sum_{e_{\rm out}\in
C_{\rm out}}P(e_{\rm out},e_{\rm in})\,,
\]
where $P(e_{\rm out},e_{\rm in})$ is the probability that an open
path from $e_{\rm in}$ to $e_{\rm out}$ exists. That is,
$\overline g$ is twice the mean number of open paths connecting
$C_{\rm in}$ to $C_{\rm out}$. \em

We describe the proof of these two Theorems using the
supersymmetric path integral method of Ref.~\refcite{JC}: the
combinatorial method in Ref.~\refcite{BCC} is perhaps more
illustrative of why the result holds, but it involves heavier
algebra.

In the standard way, the Green function $G$, being the inverse of
a matrix,  may be written as a gaussian integral over commuting
(bosonic) variables. The notation is a little cumbersome but the
basic idea is simple. Label each end of a given edge $e$ by $e_R$
and $e_L$, in the direction of propagation $e_R\to e_L$. Introduce
complex integration variables $b_R(e)$ and $b_L(e)$, each of which
is a 2-component column vector in the SU$(2)$ space. Then
\begin{equation}\label{grep}
G(e_f,e_i;z)=\langle b_L(e_f)b_L(e_i)^\dag\rangle=
\frac{\int\prod_e[db_L(e)][db_R(e)]b_L(e_f)b_L(e_i)^\dag\,e^{W[b]}}
{\int\prod_e[db_L(e)][db_R(e)]e^{W[b]}}
\end{equation}
where $W[b]=W_{\rm edge}+W_{\rm node}$ and
\[
W_{\rm edge}=z\sum_eb^\dag_L(e)U_eb_R(e)\,,\qquad W_{\rm
node}=\sum_n\sum_{ji}b^\dag_R(e_j')S_{ji}b_L(e_i)\,.
\]
We use the notation $\langle\cdots\rangle$ to denote averages with
respect to this gaussian measure. The measure for each integration
is
\[
\int[db]=(1/\pi^2)\int e^{-b^\dag b}\,d{\rm Re}\,b\,d{\rm
Im}\,b\,.
\]
For a finite graph there are a finite number of integrations and
the integral is convergent for $|z|<1$.

The next step is to average over the quenched random matrices
$U_e$. As usual this is difficult because these appear in both the
numerator and denominator of (\ref{grep}). This can be addressed
using replicas, or, much more effectively in this case, by adding
an anticommuting (fermionic) copy $(f,\bar f)$ of each pair
bosonic variables $(b,b^\dag)$. Note that each $f$ is also a
2-component column vector in SU$(2)$ space, and each $\bar f$ a
row vector. The Grassman integration over these is defined by
\[
\int[df]=\int e^{-\bar ff}\,dfd\bar f\,,
\]
where
\[
\int[df]f=\int[df]\bar f=0\quad\mbox{and}\quad
\int[df]1=\int[df]f\bar f=1\,.
\]
Integrating over these cancels the denominator in (\ref{grep}) so
that
\begin{equation}\label{srep}
G(e_f,e_i;z)=
\int\prod_e[db_L(e)][db_R(e)][df_L(e)][df_R(e)]f_L(e_f)\bar
f_L(e_i)\,e^{W[b]+W[f]}\,.
\end{equation}
The action $W[b]+W[f]$ is supersymmetric under rotating bosons
into fermions and \em vice versa\em, so we can replace the bosonic
fields at $e_i$ and $e_f$ by fermionic ones and consider $\langle
f_L(e_f)\bar f_L(e_i)\rangle$.

Now consider the average over $U$ on a given edge. This has the
form
\begin{equation}\label{singleedge}
\int dU\exp(zb_L^\dag Ub_R+z\bar f_LUf_R)\,.
\end{equation}
Because the anticommuting fields square to zero, the expansion of
the exponential in powers of the second term terminates at second
order. The gauge symmetry discussed earlier shows that the purely
bosonic zeroth order term is in fact independent of $b_L^\dag$ and
$b_R$ and is in fact unity. The second order term is also
straightforward: carefully using the anticommuting property we see
that is actually proportional to $\det U=1$, times the determinant
of the SU$(2)$ matrix of fermion bilinears with elements $\bar
f_{Li}f_{Rj}$, where $i$ and $j$ take the values 1 or 2. The first
order term then simply converts this into something
supersymmetric. The conclusion is that the integral
(\ref{singleedge}) equals $1+\ffrac12z^2\det{\bf M}$ where the
$2\times2$ matrix $\bf M$ has components
$M_{ij}=b^\dag_{Li}b_{Rj}+\bar f_{Li}f_{Rj}$. This can be
rearranged in the form
\[
1+z^2\big[(1/\sqrt2)(b^\dag_{L1}\bar f_{L2}-b^\dag_{L2}\bar
f_{L1})\big]\big[(1/\sqrt2)(b_{R1}f_{R2}-b_{R2}f_{R1})\big]
+z^2\big[\bar f_{L1}\bar f_{L2}\big]\big[f_{R2}f_{R1}\big]\,.
\]
Each expression in square brackets is an antisymmetric SU$(2)$
singlet. The three terms above describe the propagation of either
nothing, a fermion-boson ($fb$) singlet, or a fermion-fermion
($ff$) singlet along the edge $e$. This is a remarkable
simplification: before the quenched average, each edge may be
traversed many times, corresponding to the propagation of
multi-particle states. It is this which gives one of the principle
simplifications of the SU$(2)$ case, that does not happen for
U$(1)$, one of the reasons this is much more difficult.

This result shows that single fermions or bosons cannot propagate
alone, so $\overline{G(e_f,e_i)}=0$ if $e_f\not=e_i$. A non-zero
correlation function is however
\begin{equation}\label{ffcorr}
\langle f_1(e_f)f_2(e_f)\bar f_2(e_i)\bar f_1(e_i)\rangle=
\overline{G_{11}G_{12}-G_{21}G_{22}}=\overline{\det G(e_f,e_i,z)}
\end{equation}
Now $G$, being a real linear combination of products of SU$(2)$
matrices, can in fact always be written in the form
$\lambda\widetilde G$ where $\lambda$ is real and $\widetilde
G\in$SU$(2)$.\footnote{This follows from the representation
$U=\cos\alpha+i({\bf\sigma}\cdot{\bf n})\sin\alpha$.} Hence $\det
G=\lambda^2$ and $G^\dag G=\lambda^2{\bf 1}$, so ${\rm Tr}\,G^\dag
G=2\det G$. When $z=1$ this gives the point conductance between
$e_i$ and $e_f$.

When $e_i=e_f=e$, however, we can always insert a pair of fermion
fields $f_2(e)\bar f_2(e)$ into the correlator $\langle f_1(e)\bar
f_1(e)\rangle$ at no cost, so that in fact
\[
\overline{G(e,e;z)_{11}}=\overline{G(e,e;z)_{22}}=\overline{\det
G(e,e;z)}\,.
\]
Thus both the mean density of states and the mean conductance are
proportional to $\overline{\det G}$ and therefore are given by the
correlation function (\ref{ffcorr}).

The next step is to consider propagation through the nodes. Note
that we can now drop the distinction between $b_R$ and $b_L$, etc.
The contribution from a given node takes the form
\begin{equation}\label{node}
\prod_iA_{\alpha_i'}(e_i')\,{\cal
S}\,\prod_jA^\dag_{\alpha_j}(e_j)\,,
\end{equation}
where $A_1=1$, $A_2=(1/\sqrt2)(b_1f_2-b_2f_1)$, $A_3=f_1f_2$, and
\[
{\cal S}=\exp\left(\sum_{ij}b^\dag_iS_{ij}b_j+\bar
f_iS_{ij}f_j\right)\,,
\]
where $S_{ij}$ is the matrix in (\ref{cossin}). Since this
expression conserves fermion and boson number, it follows that the
total numbers of $(fb)$ and $(ff)$ singlets are also conserved.
Also, only terms second order in the $S_{ij}$ survive. In fact,
after a little algebra (\ref{node}) reduces to\footnote{In
Ref.~\refcite{JC} this was carried out for a general node of
arbitrary coordination, with excruciating algebra.}
\[
\delta_{\alpha_1'\alpha_1}\delta_{\alpha_2'\alpha_2}S_{11}S_{22} -
\delta_{\alpha_2'\alpha_1}\delta_{\alpha_1'\alpha_2}S_{21}S_{12}
=\delta_{\alpha_1'\alpha_1}\delta_{\alpha_2'\alpha_2}\cos^2\theta_n
+\delta_{\alpha_2'\alpha_1}\delta_{\alpha_1'\alpha_2}\sin^2\theta_n\,.
\]

These two terms correspond to the decomposition of the node $n$
described earlier. It shows that, on performing the quenched
average, the quantum network model is equivalent to a classical
one in which $\cal G$ is decomposed into disjoint loops, and along
each loop propagates either an $(fb)$ singlet, an $(ff)$ singlet,
or nothing. Theorems (\ref{th1},\ref{th2}) now follow
straightforwardly. We argued above that the mean diagonal Green
function $\overline{G(e,e;z)}$ is given by the $(ff)$ correlation
function $\langle[f_1(e)f_2(e)][\bar f_1(e)\bar f_2(e)]\rangle$.
So, in each decomposition of $\cal G$, there must be an $(ff)$
pair running around the unique loop containing $e$. This gets
weighted by a factor $z^{2L}$. Around all the other loops we can
have either an $(ff)$ pair, a $(bf)$ pair, or just 1. The $(ff)$
pair, being itself bosonic, gives $z^{2L}$ for a loop of length
$L$, while the $(bf)$ pair, being fermionic, gives $-z^{2L}$.
These two cancel (by supersymmetry), leaving a factor 1 for every
loop other than the one containing $e$. The argument for
Theorem~\ref{th2} for the conductance works in the same way.

Note that these methods may be extended to the quenched averages
of other observables in the quantum model, although the density of
states and the conductance are most important. However not all
quantities of interest can be treated in this fashion. For
example, the fluctuations in the conductance involve
\[
\overline{\big(G(e_{\rm out},e_{\rm in})^\dag G(e_{\rm out},e_{\rm
in})\big)^2}\,,
\]
and, in order to treat this, we would need to double the number of
degrees of freedom in the integral representation. Many of the
formulas which are special to SU$(2)$ integrations then no longer
hold. In this context it is important to note that the conductance
fluctuations in the quantum model are \em not \em given by the
fluctuations in the number of paths connecting the two contacts in
the classical model. (If this were the case, the quantum system
would be behaving completely classically!)

\begin{figure}
\centerline{\includegraphics[width=8cm]{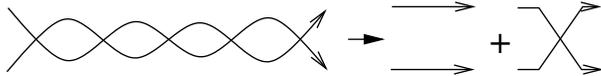}}
\caption{Graph corresponding to a single link in the Sp$(4)$
model, and its topologically distinct decompositions. Each node
corresponds to the $S$-matrix which is the second factor in
(\ref{Sp4}), so each term in the decomposition is equally
weighted.} \label{fig:Sp4}
\end{figure}

An amusing application\cite{BCC} of the general theorems is to
consider single edge $e$, closed on itself, but take $U_e\in{\rm
Sp}(N)$ with $N$ even and $>2$ in general. A general Sp$(N)$
matrix may be built up in terms of successive Sp$(2)$ rotations in
overlapping 2-dimensional subspaces. For example for Sp$(4)$ we
may write a general matrix in the form
\begin{equation}\label{Sp4}
\frac1{\sqrt2}\left(\begin{array}{cc}U_1&0\\0&U_2\end{array}\right)
\left(\begin{array}{cc}{\bf 1}&{\bf 1}\\-{\bf 1}&{\bf
1}\end{array}\right)
\end{equation}
where $U_1$ and $U_2$ are independent Sp$(2)$ matrices. If these
are drawn from the invariant measure on Sp$(2)$, then the product
of a large number of independent such matrices will converge to
the invariant measure on Sp$(4)$. Thus, in a particular Sp$(2)$
basis, $\cal G$ has the form shown in Fig.~\ref{fig:Sp4}.
After applying Theorem \ref{th1}, each decomposition corresponds
to a permutation of the different channels corresponding to the
basis chosen in (\ref{Sp4}). This generalizes to arbitrary $N$. If
we now connect opposite ends of Fig.~\ref{fig:Sp4} to make a
closed graph, we find, after the decomposition, all possible
lengths $L$ of loops from 1 to $\frac12N$, with equal
probabilities. Thus $P(L)=2/N$ for $1\leq L\leq\frac12N$, and zero
otherwise. Using (\ref{dos}), this gives for the density of
eigenvalues of a random Sp$(N)$ matrix
\[
\rho(\epsilon)=\frac{N+1}{2\pi N}\left(1-\frac{\sin(N+1)\epsilon}
{(N+1)\sin\epsilon}\right)\,,
\]
in agreement with Ref.~\refcite{Zirn}.

\section{Two-dimensional models}\label{sec:2d}
In this section we discuss the consequences of the main theorems
for specific 2d lattices relevant to physical problems.
\subsection{The L-lattice}\label{sec:Llattice}
This is the lattice, illustrated in Fig.~\ref{fig:Llattice}, used
by the original Chalker-Coddington model for the quantum Hall
plateau transition. The reasons for choosing this lattice were
discussed in Sec.~\ref{sec:intro}. In the class C version of this,
the same lattice is used, the only difference being that the
quenched random U$(1)$ phases on each edge are replaced with
SU$(2)$ matrices. We also recall that, because of the checkerboard
nature of the potential, where even plaquettes correspond to $V>0$
and odd ones to $V<0$, the angles $\theta_n$, which represent the
degree of anisotropy of the tunnelling at the nodes, are in fact
staggered: $\theta_n=\theta$ on the even sublattice, and
$(\pi/2)-\theta$ on the odd sublattice. Thus for $\theta=0$ all
the loops in the decomposition of $\cal G$ will be the minimum
size allowed, surrounding the even plaquettes, and for
$\theta=\pi/2$ they will surround the odd plaquettes. Away from
these extreme values, the loops will be larger. If there is a
single transition it must occur at $\theta=\pi/4$.

\begin{figure}
\centerline{\includegraphics[width=4.2cm]{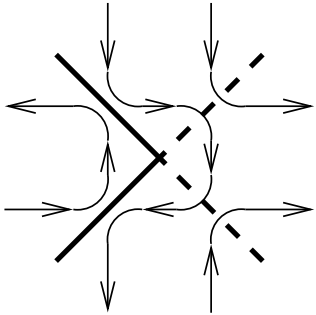}}
\caption{The mapping between decompositions of the L-lattice and
bond percolation on the lattice ${\cal L}'$. Open bonds are shown
as thick lines, closed as dashed lines.} \label{fig:perc1}
\end{figure}

The mapping to square lattice bond percolation for this model is
exact. Consider independent bond percolation on the square lattice
$\cal L'$, rotated by $45^\circ$ with respect to the original,
whose sites lie at the centers of the even plaquettes of the
original lattice. See Fig.~\ref{fig:perc1}.
Each edge of $\cal L'$ intersects a node of the original one. We
declare it to be open, with probability $p=\cos^2\theta$, or
closed, with probability $1-p=\sin^2\theta$, according to the way
the node is decomposed in the classical loop model on $\cal G$.
There is thus a 1-1 correspondence between decompositions of $\cal
G$ and bond percolation configurations on $\cal L'$. We can also
consider the dual lattice $\cal L''$ whose vertices are at the
centers of the odd plaquettes of the original lattice. Each edge
of this lattice crosses a unique edge of $\cal L'$, and we declare
it to be open, with probability $1-p$, if the corresponding edge
of $\cal L'$ is closed, and \em vice versa\em. For each
percolation configuration on $\cal L'$, there is a corresponding
dual configuration on $\cal L''$. The clusters and dual clusters
fill the plane without intersecting each other. (For this purpose
it is necessary to regard single isolated sites as clusters.)
Close to $p_c=\frac12$ many clusters nest inside dual clusters and
\em vice versa\em.

For a given decomposition of $\cal G$, the loops give the \em
hulls \em of percolation clusters on $\cal L'$ and dual clusters
on $\cal L''$. For a closed simply connected lattice $\cal G$,
these are closed curves which simultaneously circumscribe a
cluster and inscribe a dual cluster, or \em vice versa\em. For
open boundary conditions, some of these curves may begin and end
on the boundary.

From this mapping to 2d percolation, and known exact and
conjectured results about the latter, many results about the
SU$(2)$ network model on the L-lattice may be deduced\cite{GRL}.
At the transition, the conductance between two bulk points a
distance $r$ apart is given by the probability that they are on
the same loop, which is known to decay as $|r|^{-2x_1}$ where
$x_1=\frac14$ \cite{1leg}. The conductance of a rectangular sample
with contacts along opposite edges is given by the mean number of
hulls which cross the sample. At the critical point this depends
only its aspect ratio $L_1/L_2$, in a complicated but calculable
way\cite{JCcond}. For $L_1/L_2\gg1$ (where $L_1$ is the length of
the contacts) it goes like $\bar g(L_1/L_2)$ where $\bar g$ is the
universal critical conductance which, from conformal field theory
results applied to percolation, takes the value
$\sqrt3/2$.\cite{JCcond} In Sec.~\ref{sec:proof} we showed that
the mean density of states is given by the probability $P(L)$ that
a given edge is on a loop of length $L$. At criticality, this
probability scales like $L^{-x_1}$ where $d_f=2-x_1=\frac74$ is
the fractal dimension of percolation hulls.\cite{1leg}. Also
$|z-1|\propto|E|$, which is conjugate to $L$, has RG eigenvalue
$y_1=2-x_1$. This means that the singular part of the mean density
of states behaves like\cite{GRL}
\[
\overline{\rho(E)}\sim|E|^{x_1/y_1}=|E|^{1/7}\,.
\]
A number of other exponents, including the usual percolation
correlation length exponent $\nu=\frac43$, were identified in the
physics of the spin quantum Hall transition in Ref.~\refcite{GRL}.

\subsection{The Manhattan lattice}\label{sec:Manhattan}
\begin{figure}
\centerline{\includegraphics[width=5cm]{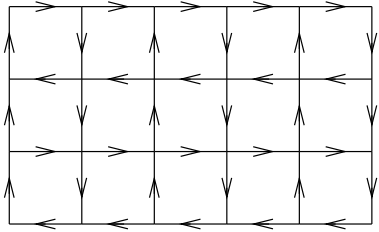}}
\caption{The Manhattan lattice.} \label{fig:manhattan}
\end{figure}

Although the $L$-lattice is the natural candidate for studying the
spin quantum Hall transition, the mapping discussed in
Sec.~\ref{sec:proof} is of course valid for any orientation of the
edges of a square lattice, and one can legitimately ask whether
other possibilities lead to interesting physics. The expectation,
based on the continuum classification of localization universality
classes, is that unless there is some special symmetry, such as
occurs for the L-lattice with its sublattice symmetry, all states
in 2d will be exponentially localized and therefore almost all
loops in the classical model will have finite length. This was
studied for the Manhattan lattice in Ref.~\refcite{BCO}. On this
lattice all edges in the same row or column are oriented in the
same direction, and these alternate, see Fig.~\ref{fig:manhattan}.
\begin{figure}
\centerline{\includegraphics[width=5.5cm]{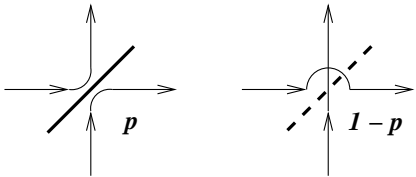}}
\caption{The decomposition of each node of the Manhattan lattice
and its relation to percolation on ${\cal L}'$. Open bonds shown
as solid thick lines, closed as dashed.}
\label{fig:manhattandecomp}
\end{figure}
This resembles the one-way system of streets and avenues in
Manhattan. At each corner, the driver can go straight on, or turn
either left or right according to the parity of the intersection.

Consider an SU$(2)$ network model on this lattice. The
decomposition of the lattice corresponds to replacing each node
either by a crossing, with probability $1-p$, say, or an
avoidance, with probability $p$ (see
Fig.~\ref{fig:manhattandecomp}). Note that in this case the loops
on the decomposed lattice are in general non-planar. Nevertheless
it is possible to make rigorous progress using a mapping to
percolation, owing to the sublattice structure. Consider once
again bond percolation on the $45^\circ$-rotated lattice $\cal L'$
(see Fig.~\ref{fig:manhattandecomp}). An edge is declared open if
the corresponding node is decomposed in such a way that the paths
turn by $90^\circ$, as shown. The open edges once again form
connected clusters which enclose, and are enclosed by, percolation
hulls, and similarly for the dual clusters. A little reflection
shows that each loop of the decomposed Manhattan lattice is
constrained to lie on or between neighboring hulls which enclose a
dual cluster. For $p>p_c=\frac12$ the dual clusters are almost
surely finite, and therefore so are their surrounding hulls, and
therefore also the loops of the decomposed $\cal G$. Therefore for
$p>\frac12$ the SU$(2)$ network model on the Manhattan lattice is
in the localized phase. An unproven conjecture, consistent with
our expectations for Anderson localization in 2d, is that this
happens for all $p>0$. Simulations\cite{BCO} of the classical loop
model indicate that this is the case for $p>0.2$. The field theory
arguments to be discussed later indicate that there should always
be a finite localization length, diverging as
$\xi\sim\exp\big({\rm const.}/p^3\big)$ as $p\to0$.

\subsection{Other 2d lattices}
We may consider other orientations of the square lattice. In
general a given orientation corresponds to a configuration of the
six-vertex model, which satisfies the `ice rule' that there are
two incoming and two outgoing arrows at each vertex, or node. The
number of such allowed configurations grows exponentially with the
size of the lattice, but we could, for example, consider a
randomly oriented lattice in which the weights for different types
of node are given by the 6-vertex model. In that case the
L-lattice and Manhattan lattice are just particular extreme points
of the parameter space. Once again we can associate an edge of the
lattice $\cal L'$ or $\cal L''$ with the decomposition of a node
where the path has to turn. These edges can be thought of as
two-sided mirrors, reflecting all the paths which impinge on
either side. The study of these `mirror models' as models for
classical localization has been extensive (see, for example
Ref.~\refcite{mirrormodels}, and references quoted therein),
although on the whole only simulational results are available.
However it is important to realize that arbitrary mirror models do
not in general lead to a unique orientation for the edges: a path
may traverse a given edge in both directions. Thus the set of
mirror configurations corresponding to quantum network models is a
subset, and it would be wrong to infer general conclusions about
these from the study of the wider problem. Nevertheless, one
expects that, for a sufficiently high density of randomly oriented
mirrors, the paths are finite and so the states are localized. An
interesting and unresolved question, however, is what happens at
low mirror density, when the mean free path is large. Expectations
from quantum localization would then suggest that the paths in
such models are still localized on large enough scales, unless
there is some special symmetry like that of the L-lattice.

\section{Three-dimensional models}\label{sec:3d}
We now discuss some results for class C network models on 3d
lattices. It is possible, of course, to consider layered 2d
lattices which might be used as models for quantum Hall physics in
multilayered systems. For example, in a bilayer system consisting
of two coupled L-lattices, depending on the strength of the
coupling between the layers, one expects to see either two
separate transitions between states of Hall conductance $0$, $1$
and $2$. These can be simply understood in terms of the classical
model and an equivalent percolation problem.\cite{Bthesis}
However, such models do not capture some of the important
properties of real bilayer systems which depend on
electron-electron interactions.

One motivation for studying truly 3d class C network models is to
shed light on the physics of the localization transition in 3d. So
far, these have been carried out either numerically or by mapping
to 3d percolation. However the restriction to four-fold
coordination in order that the equivalent classical model has
non-negative weights\cite{JC} means that it is necessary to use
loose-packed lattices.

\subsection{Diamond lattice}\label{sec:diamond} 
\begin{figure}
\centerline{\includegraphics[width=10cm]{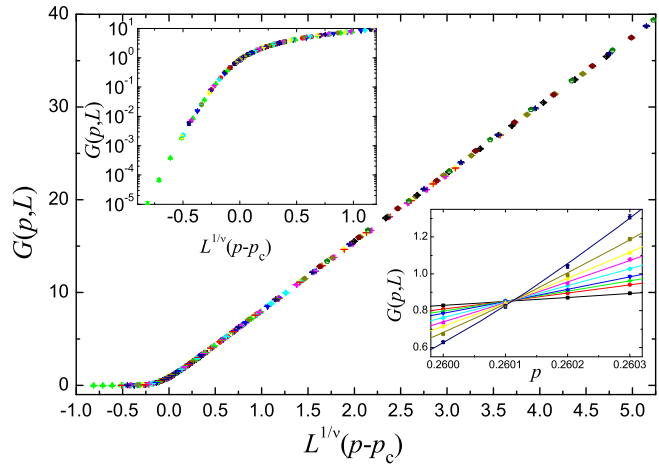}}
\caption{Conductance as a function of $(p-p_c)L^{1/\nu}$,
illustrating scaling collapse. Upper inset: same data on a
logarithmic scale. Lower inset: conductance as a function of $p$
for several values of $L$. Lines are a scaling fit described in
detail in Ref.~\refcite{OSC}. (Reproduced from Ref.~\refcite{OSC}
with permission of the authors.)} \label{fig:diamond}
\end{figure}

The most extensive
numerical simulations have been carried out on the diamond
lattice.\cite{OSC} Although this lattice has cubic symmetry,
assigning the orientations of the edges breaks this down to
tetragonal, inducing an anisotropy. However, numerical tests show
that this is not very great, giving, for example, a ratio of about
$1.1$ between the conductances in the two distinct directions in
the conducting phase. As expected, the model exhibits a sharp
transition at $p=p_c$ between an insulating phase and a conducting
phase. This is shown, for example, in data for the conductance
$G(p,L)$ of a cubic sample of linear size $L$, which, according to
the Theorem \ref{th2}, is given by the mean number of open paths
between two opposite faces. In the insulating phase $p<p_c$ this
should approach zero as $L\to\infty$, while for $p>p_c$ we expect
ohmic behavior with $G(p,L)\sim\sigma(p)L$. Close the critical
points we expect finite-size scaling of the form
\[
G(p,L)=f\big(L/\xi(p)\big)\,,
\]
where the localization length $\xi(p)\sim |p-p_c|^{-\nu}$. Thus
the data should show collapse when plotted as a function of
$(p-p_c)L^{1/\nu}$, and this is clearly exhibited in
Fig.~\ref{fig:diamond}.
The best fitted value for $\nu$, taking into account corrections
to scaling, is\cite{OSC} $\nu=0.9985\pm 0.0015$. The closeness of
this value to unity, the value predicted by a first-order result
$\nu^{-1}=\epsilon+O(\epsilon^2)$ of the $2+\epsilon$-expansion
(see Sec.~\ref{sec:fieldtheory} and Ref.~\refcite{senthil}) is
remarkable, but perhaps a coincidence.

At $p=p_c$, the weighted number of return paths of length $L$,
behaves as $P(L)\sim L^{-x_1/y_1}$, as discussed in
Sec.~\ref{sec:Llattice}. In Ref.~\refcite{OSC} this exponent is
denoted by $2-\tau$, where numerically $\tau=2.184\pm0.003$. This
exponent is related to the fractal dimension $d_f=y_1=3-x_1$ of
the paths at $p_c$, giving $d_f=2.534\pm0.009$. By the same
arguments as in Sec.~\ref{sec:Llattice}, this gives, for example,
the singular part of the density of states
$\rho(E)\sim|E|^{\tau-2}$. Note that, because of the mapping to
the classical problem for which far larger systems can be studied,
the error bars on these exponents are much smaller than those
quoted for the conventional 3d Anderson transition.

\subsection{3d L-lattice and Manhattan lattice}\label{3dL}
It is possible to construct 3d oriented lattices with cubic
symmetry which are direct analogues of the 2d L-lattice and
Manhattan lattices discussed in Sec.~\ref{sec:2d} and for which
the arguments relating the classical models to percolation can be
generalized.

\begin{figure}
\centerline{\includegraphics[width=4.5cm]{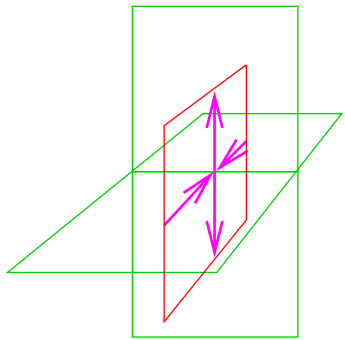}}
\caption{A node of $\cal G$ (purple) is formed by the intersection
of four faces of ${\cal C}_2$ (green) and one face of ${\cal C}_1$
(red), and \em vice versa\em. The orientation shown corresponds to
the 3d L-lattice.} \label{fig:node}
\end{figure}

Consider two interpenetrating cubic lattices ${{\cal
C}}_1\equiv{\bf Z}^3$ and ${{\cal C}}_2\equiv({\bf Z}+\frac12)^3$.
Each face of ${\cal C}_1$ intersects an edge of ${\cal C}_2$ at
its midpoint. The four faces of ${\cal C}_2$ which meet along this
edge intersect the given face of ${\cal C}_1$ along two mutually
perpendicular lines, also perpendicular to the edge (see
Fig.~\ref{fig:node}). These lines form part of the lattice $\cal
G$. The same is true, interchanging the roles of ${\cal C}_1$ and
${\cal C}_2$. The full graph $\cal G$ is the lattice formed by the
intersection of the faces of ${\cal C}_1$ with those of ${\cal
C}_2$. The nodes of $\cal G$ lie on the midpoints of the edges of
${\cal C}_1$ (the centers of the faces of ${\cal C}_2$) and \em
vice versa\em, and have coordination number 4. Clearly $\cal G$
has cubic symmetry.

For the L-lattice on $\cal G$, the orientation of the edges is
chosen so that each node looks like the nodes of the 2d L-lattice,
as in Fig.~\ref{fig:node}. There is an overall 2-fold degeneracy
in assigning these, but once the orientation at one node is fixed,
so are the rest. The $S$-matrices of the network model, and the
corresponding probabilities $p$ and $1-p$ for the decomposition of
$\cal G$ are assigned consistent with the percolation mapping now
to be described.

\begin{figure}
\begin{center}
  \parbox{2.1in}{\includegraphics[width=3.5cm]{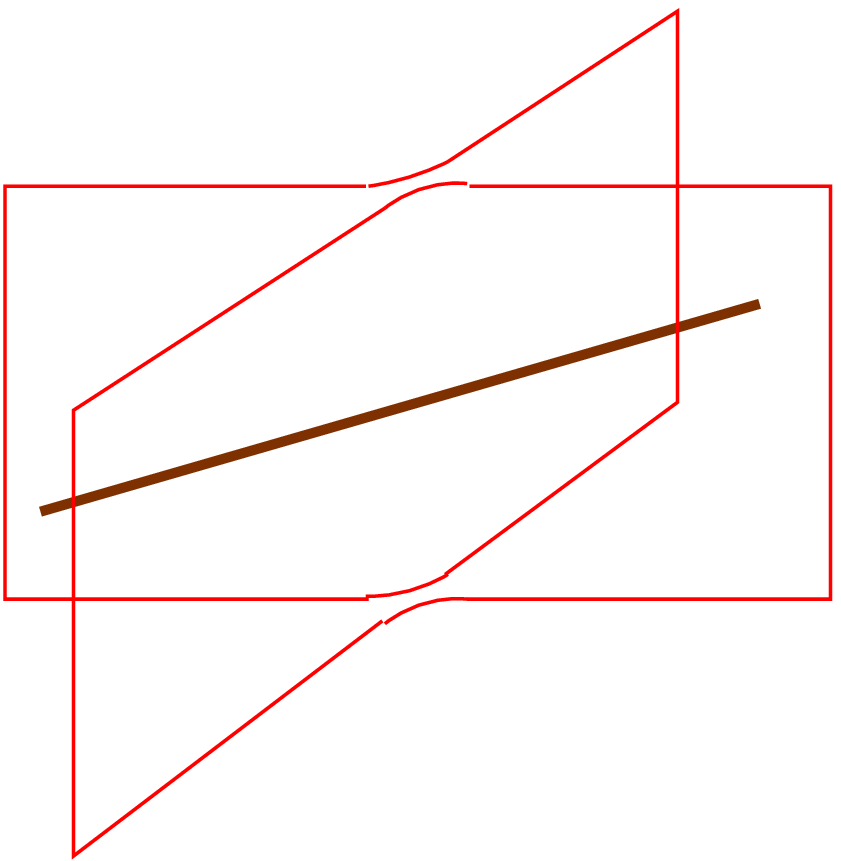}}
  \hspace*{4pt}
  \parbox{2.1in}{\includegraphics[width=3.5cm]{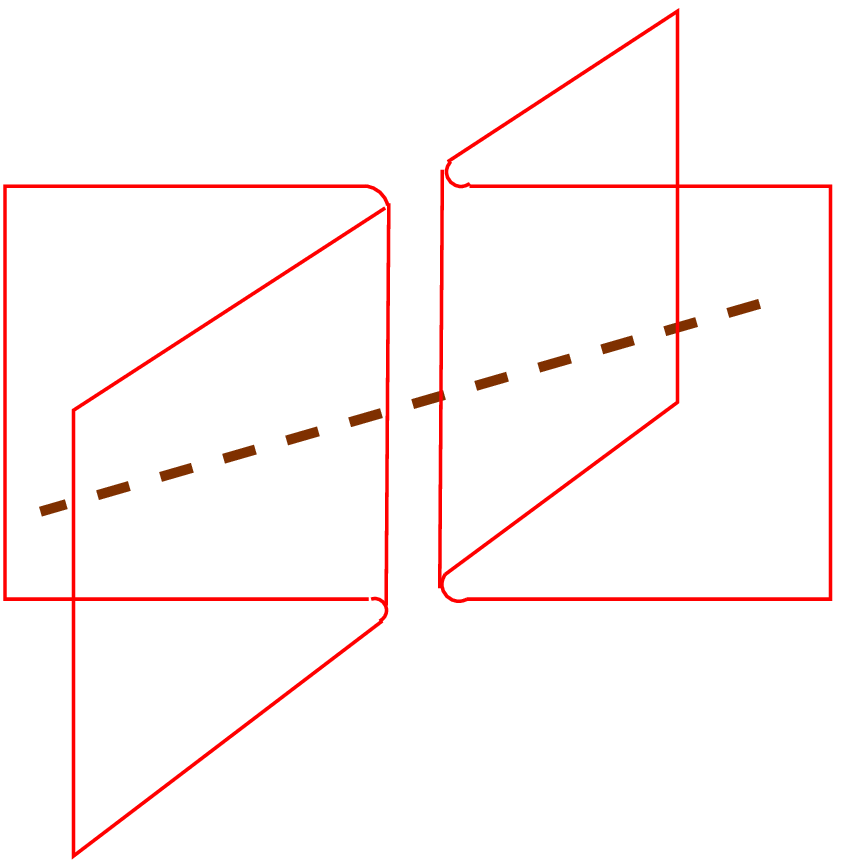}}
  \caption{Decomposition of the four faces (in red) at an edge of ${\cal C}_1$ corresponding
to bond percolation (bonds shown in brown) on an fcc sublattice of
${\cal C}_2$.} \label{fig:intersection}
\end{center}
\end{figure}

The sites of ${\cal C}_1$ may be assigned to even and odd
sublattices ${\cal C}_1'$ and ${\cal C}_1''$ according to whether
the sum of the coordinates is even or odd. Each of these lattices
is in fact a face-centered cubic (fcc) lattice. Now consider
nearest neighbor bond percolation on ${\cal C}_1'$. Each nearest
neighbor edge of ${\cal C}_1'$ intersects the midpoint of an edge
of ${\cal C}_2$, along which 4 faces of ${\cal C}_2$ intersect. A
decomposition of this edge consists in connecting up these faces
in neighboring pairs, as in Fig.~\ref{fig:intersection}.

For each edge there are two possible decompositions, and we can do
this in such a way that if the corresponding edge of the
percolation problem on ${\cal C}_1'$ is open, it passes between
the connecting pairs (see Fig.~\ref{fig:intersection}), just as in
Fig.~\ref{fig:perc1} in 2d. If on the other hand the edge is
closed, then it intersects both pairs. Equivalently, we can
consider percolation on the `dual' fcc lattice ${\cal C}_1''$.
Each edge of this lattice intersects one edge of ${\cal C}_1'$ at
the midpoint of an edge of ${\cal C}_2$. We declare the edge of
${\cal C}_1''$ to be open if the intersecting edge of ${\cal
C}_1'$ is closed, and \em vice versa\em.

For a finite lattice, each decomposition of the edges of ${\cal
C}_2$ divides the faces of ${\cal C}_2$ into a union of dense,
non-intersecting, closed surfaces, in the same way that a
decomposition of the corresponding nodes of the 2d square lattice
divides the edges into non-intersecting closed loops. These closed
surfaces form the hulls of the bond percolation clusters on the
fcc lattice ${\cal C}_1'$ and its dual ${\cal C}_1''$. That is,
each closed surface either touches a unique cluster externally and
a unique dual cluster internally, or \em vice versa\em.

However, this is only half the description. The edges of $\cal G$
are formed by the intersection of the faces of ${\cal C}_1$ with
those of ${\cal C}_2$. Therefore we need to also decompose the
faces of ${\cal C}_1$. This is carried consistent with another,
independent, percolation problem on an fcc sublattice ${\cal
C}_2'$ of ${\cal C}_2$, and its dual ${\cal C}_2''$. To each
double decomposition
\begin{figure}
\centerline{\includegraphics[width=6cm]{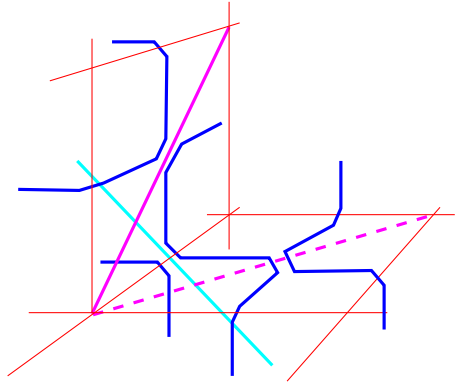}} \caption{A
decomposition of $\cal G$ on the 3d L-lattice, showing how the
loops (blue) reflect off the open bonds of percolation on the fcc
lattices ${\cal C}_1'$ (solid magenta lines) and ${\cal C}_2'$
(solid turquoise lines).} \label{fig:3dloops}
\end{figure}
of the faces of ${\cal C}_1$ into closed non-intersecting
surfaces, and similarly of the faces of ${\cal C}_2$, corresponds
a unique decomposition of $\cal G$ into closed loops.

This is the classical loop model which we seek. Unlike the 2d
case, it corresponds to \em two \em independent percolation
problems, one on an fcc lattice of ${\cal C}_1$, the other on an
fcc sublattice of ${\cal C}_2$. Each closed loop of $\cal G$ is
formed by the intersection of a closed surface made up of faces of
${\cal C}_1$ and a closed surface made up of faces of ${\cal
C}_2$. (It is of course possible that such pairs of closed
surfaces intersect in more than one loop of $\cal G$, or not at
all.) It alternately `reflects' off an open edge of percolation on
${\cal C}_1'$ (or an open dual edge of ${\cal C}_1''$), then an
open edge of ${\cal C}_2'$ or of ${\cal C}_2''$, and so on (see
Fig.~\ref{fig:3dloops}).

We now discuss the physics of this model and its implications for
the class C network model. In principle we can assign different
probabilities, $p_1$ and $p_2$, to the two independent percolation
problems on ${\cal C}_1$ and ${\cal C}_2$. The phase diagram is
symmetric under the duality symmetry $p_j\to1-p_j$, so we can
restrict attention to the quadrant $p_1\leq\frac12$,
$p_2\leq\frac12$. For either $p_1$ or $p_2$ less than the bond
percolation threshold $p^{\rm fcc}_c\approx0.12$ \cite{ziff},
since the clusters are finite so are the hulls on either ${\cal
C}_1$ or ${\cal C}_2$ (or both), and, since the loops on $\cal G$
are formed by their intersection, these must be finite also,
corresponding to a localized phase. Preliminary simulations of the
model\cite{Ikhlef} for values of $p_j$ close to $\frac12$ appear
to show that the loops are no longer finite, but, unlike the case
of the diamond lattice\cite{OSC}, neither are they simple random
walks on large scales with fractal dimension 2, as would be
expected of a sample exhibiting ohmic behavior. In fact, their
fractal dimension appears to be close to 3, indicating that they
are space filling. This may be a pathology of this model. It is
known that the fractal dimension of cluster boundaries for $p>p_c$
is 3 -- that is, a finite fraction of the infinite cluster, which
has dimension 3 because it contains a finite fraction of the
sites, is on its boundary. The loops for $p>p_c$ are formed by the
intersection of random such $d_f=3$ objects, so it is perhaps not
surprising that they should also have $d_f=3$.

For this reason it may be more useful to consider a 3d version of
the Manhattan lattice. This is based on the same graph $\cal G$,
but the edges are oriented so that along each line through the
lattice they point in the same direction, and the lines alternate
direction transversally. As in the 2d case, the probability of
turning at a given node is $p$ and of going straight on $1-p$. (In
principle we could again take different values of $p$ at the two
types of nodes corresponding to ${\cal C}_1$ and ${\cal C}_2$.)
Analogously with the 2d case (see Sec.~\ref{sec:Manhattan}), we
can introduce an associated percolation problem on fcc sublattices
of ${\cal C}_2$ and ${\cal C}_1$, so that if, in the decomposition
of $\cal G$, the paths turn at a given node, they reflect off an
open edge. This means that, as in 2d, they are constrained to lie
on regions occupied by the dual clusters of each percolation
problem. Thus, for $p>1-p^{\rm fcc}_c\approx0.88$, the loops of
$\cal G$ are almost surely finite in length, corresponding to the
existence of a localized phase of the corresponding class C
network model.

On the other hand, for small $p$ we expect a finite fraction of
the paths to escape to infinity on a infinite lattice,
corresponding to extended states. In this case the particle
following a path will, almost, all the time, go straight ahead,
with only a small probability $p\ll1$ of turning. In this case,
the whether the particle has traversed an even or odd number of
edges should be unimportant, leading to an effective simple random
walk with diffusion constant $O(p^{-1})$ on intermediate distance
scales. On larger scales, the walk may revisit regions it has in
the past, but, unlike the case of 2d, this is unlikely because the
3d random walk is not recurrent. However, as appealing as this
argument may be, it is not rigorous, and indeed its proof appears
to be of the same order of difficulty as showing that the `true'
self-avoiding walk in 3d is asymptotically gaussian\cite{truerw,
Peliti}. Thus, at this stage, a proof of the existence of extended
states in this class of network models remains elusive.

\section{Summary and further remarks}\label{sec:further}
We have shown how the quantum $\to$ classical mapping for class C
network models helps to gain insight into the nature of Anderson
localization in both two and three dimensions. These models have
direct physical relevance in systems where time-reversal symmetry
is broken but spin rotational symmetry is preserved. The classical
models correspond to deterministic motion in a random medium with
two-sided mirrors, or, equivalently, certain kinds of
history-dependent random walks. On some lattices, including the
important examples of the L and Manhattan lattices, the paths are
associated with the hulls of percolation clusters, and rigorous
information can be inferred on whether the corresponding quantum
model is in a localized or extended phase.

However, there are a number of unresolved questions. Although the
diamond lattice and 3d Manhattan lattice discussed in
Sec.~\ref{sec:3d} are expected to exhibit an Anderson transition,
there is as yet no proof of the existence of an extended phase in
which the motion is asymptotically diffusive, although this is
strongly indicated on both numerical and other grounds. The
relation of these models to other types of history-dependent
random walks, such as the `true' self-avoiding walk, in which the
walk avoids regions it has visited in the past, is also unclear.
Like Anderson localization, $d=2$ is a critical dimension for the
true self-avoiding walk. However in this case the RG
flows\cite{Peliti,OP} are to free random walks for $d\geq2$, and
to a non-trivial stable fixed point for $d<2$, while for Anderson
localization we expect to find a non-trivial unstable fixed point
for $d>2$. This suggests that the two problems are related by a
change of sign of the interaction. However the analysis of Peliti
and Obukhov\cite{OP} shows that for history-dependent random walks
there are in fact three coupling constants which are potentially
important near $d=2$. An attempt to fit the walks on the 2d
Manhattan lattice into this picture was made in
Ref.~\refcite{BCO}. However this was not systematic and further
work needs to be done in this direction. It should be noted,
however, that a sigma-model analysis of the original class C
quantum model does give the expected unstable fixed point for
$d>2$.\cite{senthil}. A related question is that of the \em upper
\em critical dimension for the transition in this model. The
relation to interacting random walks suggests that this might be
$d=4$, as for ordinary polymers, but in this case the interactions
are not simply repulsive, so this conclusion may not hold.

\section*{Acknowledgements}
I am especially grateful to John Chalker for many informative
discussions of this subject over the years, as well as Ilya
Gruzberg, Yacine Ikhlef, Andreas Ludwig, Adam Nahum, Aleks
Owczarek, Nick Read, Tom Spencer, Bob Ziff and Martin Zirnbauer.
This work was supported in part by EPSRC Grant EP/D050952/1.


\begin{thebibliography}{9}
%
\bibitem{ChalkerCodd} J.~T.~Chalker and P.~D.~Coddington,
Percolation, tunnelling and the integer quantum Hall effect,
\emph{J. Phys. C}. {\bf 21}, 2665-2679, (1988).
%
\bibitem{PrangeJoynt} R.~E.~Prange and R.~Joynt,
Conduction in a strong field in two dimensions: The quantum Hall
effect, \emph{Phys. Rev. B}. {\bf 25}, 2943-2946, (1982).
%
\bibitem{Trugman} S.~A.~Trugman,
Localization, percolation, and the quantum Hall effect,
\emph{Phys. Rev. B}. {\bf 27}, 7539-7546, (1983).
%
\bibitem{fieldtheories} A.~M.~M.~Pruisken,
On localization in the theory of the quantized hall effect: A
two-dimensional realization of the $\theta$-vacuum, \emph{Nucl.
Phys. B}. {\bf 235}, 277-298, (1984); I. Affleck, Critical
behaviour of SU$(n)$ quantum chains and topological non-linear
$\sigma$-models, \emph{Nucl. Phys. B}. {\bf 305}, 582-596, (1988);
A.~W.~W.~Ludwig, M.~P.~A.~Fisher, R.~Shankar, and G.~Grinstein,
Integer quantum Hall transition: An alternative approach and exact
results, \emph{Phys. Rev. B}. {\bf 50}, 7526-7552, (1994);
M.~R.~Zirnbauer, Conformal field theory of the integer quantum
Hall plateau transition, arXiv:hep-th/9905054 (unpublished).
%
\bibitem{spinHall} T. Senthil,
J.~B.~Marston and M.~P.~A.~Fisher, Spin quantum Hall effect in
unconventional superconductors, \emph{Phys. Rev. B}. {\bf 60},
4245-4254, (1999).
%
\bibitem{AZ} A.~Altland and M.~R.~Zirnbauer,
Nonstandard symmetry classes in mesoscopic normal-superconducting
hybrid structures \emph{Phys. Rev. B}. {\bf 55}, 1142-1161,
(1997); M.~R.~Zirnbauer, Riemannian symmetric superspaces and
their origin in random-matrix theory, \emph{J. Math. Phys.}. {\bf
37}, 4986-5018, (1996).
%
\bibitem{ClassCnum} V.~Kagalovsky, B.~Horovitz, Y.~Avishai,
Landau-level mixing and spin degeneracy in the quantum Hall
effect, \emph{Phys. Rev. B}. {\bf 55}, 7761-7770, (1997);
V.~Kagalovsky, B.~Horovitz, Y.~Avishai and J.~T.~Chalker, Quantum
Hall Plateau Transitions in Disordered Superconductors,
\emph{Phys. Rev. Lett}. {\bf 82}, 3516-3519, (1999).
%
\bibitem{GRL} I.~A.~Gruzberg, A.~W.~W.~Ludwig and N.~Read,
Exact Exponents for the Spin Quantum Hall Transition, \emph{Phys.
Rev. Lett}. {\bf 82}, 4524-4527, (1999).
%
\bibitem{SLE} S.~Smirnov and W.~Werner,
Critical exponents for two-dimensional percolation, \emph{Math.
Res. Lett}. {\bf 8}, 729-744, (2001).
%
\bibitem{BCC} E.~J.~Beamond, J.~Cardy and J.~T.~Chalker,
Quantum and Classical Localisation, the spin Quantum Hall effect
and generalisations, \emph{Phys. Rev. B}. {\bf 65}, 214301-214310,
(2002).
%
\bibitem{JC} J.~Cardy, Network Models in Class C on Arbitrary Graphs,
\emph{Comm. Math. Phys}. {\bf 258}, 87, (2003).
%
\bibitem{Zirn} M.~R.~Zirnbauer, Supersymmetry for systems with unitary disorder:
circular ensembles, \emph{J. Phys. A}. {\bf 29}, 7113-7136,
(1996).
%
\bibitem{1leg} B.~Nienhuis, Critical behavior of
two-dimensional spin models and charge asymmetry in the Coulomb
gas, \emph{J. Stat. Phys}. {\bf 34}, 731-761, (1984).
%
\bibitem{JCcond} J.~Cardy, Linking numbers for self-avoiding loops and percolation:
application to the spin quantum Hall transition, \emph{Phys. Rev.
Lett.} {\bf 84}, 3507-3510, (2000).
%
\bibitem{BCO} E.~J.~Beamond, J.~Cardy and A.~L.~Owczarek,
Quantum and Classical Localisation and the Manhattan Lattice,
\emph{J. Phys. A}. {\bf 36}, 10251, (2003).
%
\bibitem{mirrormodels} M.~S.~Cao and E.~G.~D.~Cohen,
Scaling of particle trajectories on a lattice, \emph{J. Stat.
Phys.} {\bf 87}, 147-178, (1997).
%
\bibitem{Bthesis} E.~A.~Beamond, Ph.D. thesis (unpublished).
%
\bibitem{OSC} M.~Ortu\~no, A.~M.~Somoza and J.~T.~Chalker,
Random Walks and Anderson Localization in a Three-Dimensional
Class C Network Model, \emph{Phys. Rev. Lett}. {\bf 102},
070603-070606, (2009).
%
\bibitem{ziff} C.~D.~Lorenz and R.~M.~Ziff,
Precise determination of the bond percolation thresholds and
finite-size scaling corrections for the sc, fcc, and bcc lattices,
\emph{Phys. Rev. E.} {\bf 57}, 230–236, (1998).
%
\bibitem{Ikhlef} Y.~Ikhlef, private communication.
%
\bibitem{truerw} R.~T.~Durrett and L.~C.~Rogers, Asymptotic
Behaviour of Brownian Polymers, \emph{Probab. Theory Related
Fields} {\bf 92}, 337-349, (1992.)
%
\bibitem{Peliti} D.~J.~Amit, G.~Parisi and L.~Peliti, The
Asymptotic Behaviour of the `True' Self-Avoiding Walk, \emph{Phys.
Rev. B}. {\bf 27}, 1635–1645, (1983).
%
\bibitem{OP} S.~P.~Obukhov and L.~Peliti, Renormalisation of the
`true' self-avoiding walk, \emph{J. Phys. A}. {\bf 16}, L147-L151,
(1983.)
%
\bibitem{senthil} T.~Senthil, M.~P.~A.~Fisher, L.~Balents and
C.~Nayak, Quasiparticle Transport and Localization in High-$T_c$
Superconductors, \emph{Phys. Rev. Lett}. {\bf 81}, 4704–4707,
(1998); T.~Senthil and M.~P.~A.~Fisher, Quasiparticle density of
states in dirty high-$T_c$ superconductors, \emph{Phys. Rev. B}.
{\bf 60}, 6893–6900, (1999).
%
\end{thebibliography}
\end{document}